\documentclass{appolb}
\usepackage{graphicx,float,wrapfig,subfigure}
\usepackage{amsfonts,amsmath,amssymb,amstext}
\usepackage{latexsym}
 \usepackage{bm}
\usepackage{color}
\usepackage[normalem]{ulem}
\usepackage{placeins}
\usepackage{cite}
\usepackage{hyperref}

\begin{document}
\title{The role of thermal photons in a magnetized plasma and their significance in heavy ion collisions
\thanks{Presented at Excited QCD 2026 Workshop, Granada, Spain, 8-14 January, 2026.}}

\author{Ouyang Zhaolu, Xinyang Wang
\address{Center for Fundamental Physics, Anhui University of Science and Technology}}
\maketitle
\begin{abstract}
In this contribution, we present thermal photon production mechanisms within a magnetized quark–gluon plasma, utilizing the framework of Landau-level quantization. We examine the specific influences of the magnetic field, chemical potential, and chiral chemical potential on photon yield and polarization. By providing a more comprehensive theoretical description of photon production across the full evolution of heavy-ion collisions, our study offers a promising new avenue for resolving the "photon $v_2$ puzzle" and potentially detecting the signature of magnetic fields in heavy-ion collision experiments.
\end{abstract}
  
\section{Introduction}
Electromagnetic probes play a pivotal role in relativistic heavy-ion collisions. Due to their weak final-state interactions, they escape the medium with minimal distortion, effectively serving as an unobstructed "thermometer" of the evolving system. Consequently, thermal photons encode vital information regarding the trajectory of hot and dense QCD matter—from the primordial quark-gluon plasma (QGP) stage through to the late hadronic phase ~\cite{Turbide2008,Shen2016,Paquet2016}.

Despite their utility, a significant discrepancy remains: the direct-photon $v_2$ puzzle. Standard models suggest that because a large fraction of thermal photons is emitted during the early stages—before collective flow is fully established—their elliptic flow ($v_2$) should be relatively small. However, measurements at both RHIC and the LHC have revealed a surprisingly sizable direct-photon elliptic flow ~\cite{PHENIX2012,PHENIX2016,ALICE2019}. This persistent gap between theory and observation necessitates the exploration of additional production mechanisms. In noncentral collisions, the presence of intense magnetic fields can fundamentally alter QCD matter properties, potentially enhancing photon emission and introducing new sources of anisotropy.

In this contribution, we will present thermal photon production within a magnetized plasma, focusing on anisotropy and polarization effects and their relevance to resolving the $v_2$ puzzle. We further examine the impact of finite chemical potential and chiral chemical potential, proposing a novel methodology for detecting definitive magnetic field signatures in heavy-ion experiments ~\cite{Wang2020,WangShovkovy2021PRD,WangShovkovy2021EPJC,WangShovkovy2024,WangShovkovy2024Circular}.

\section{Thermal photons from magnetized plasma}
\begin{figure}[htb]
\centering
  \subfigure[]{\includegraphics[width=0.25\textwidth]{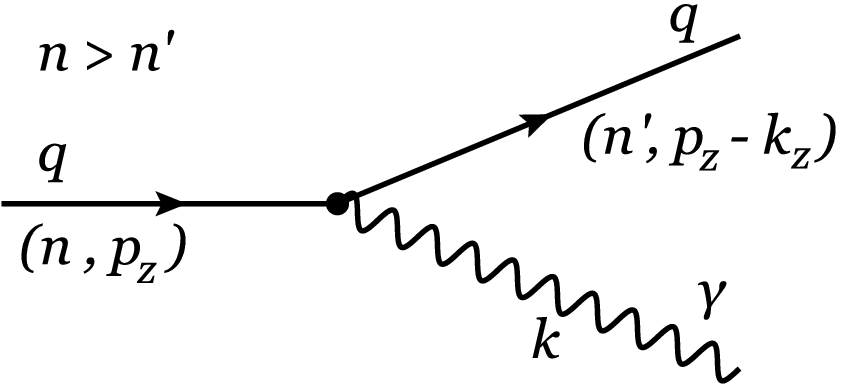}}
  \hspace{0.1\textwidth}
  \subfigure[]{\includegraphics[width=0.25\textwidth]{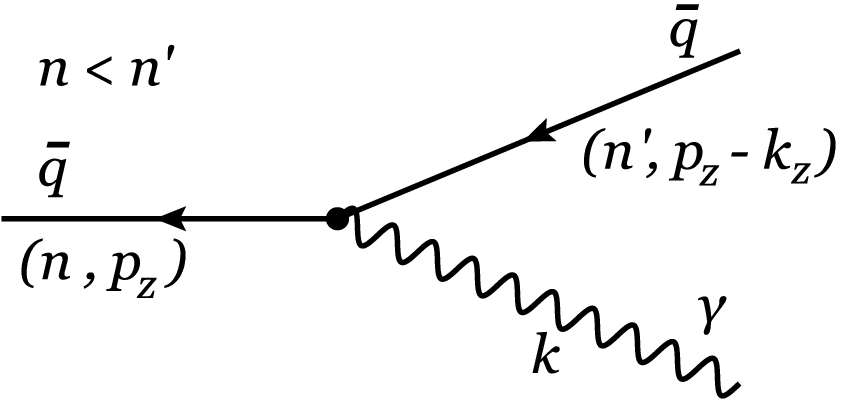}}
  \hspace{0.1\textwidth}
  \subfigure[]{\includegraphics[width=0.25\textwidth]{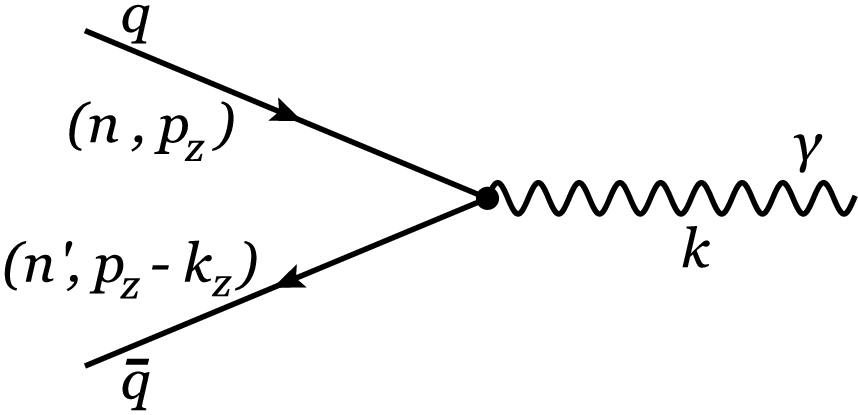}}\\
\caption{Three types of processes involving fermion states with the Landau-level indices $n$ and $n^{\prime}$:
(a) $q\to q+\gamma$, (b) $\bar{q}\to \bar{q}+\gamma$, (c) $q+ \bar{q}\to \gamma$. }
\label{LLprocesses}
\end{figure}

In a hot QCD plasma under a background magnetic field, the transverse motion of charged fermions is quantized into Landau levels. As a result, the photon emission processes are qualitatively modified compared with the zero-field case. The relevant channels include quark and antiquark synchrotron-like radiation, $q \to q + \gamma$ and $\bar q \to \bar q + \gamma$, as well as quark-antiquark annihilation, $q+\bar q \to \gamma$, accompanied by transitions between different Landau levels. These processes are schematically illustrated in Fig.~\ref{LLprocesses}.

The thermal photon production rate can be obtained from the absorptive part of the retarded photon polarization tensor in the magnetized medium, 
\begin{eqnarray}
	 \frac{d^3R}{k_T d k_T d \phi d y} = -\frac{1}{(2\pi)^3}\frac{\mbox{Im}\left[\Pi^{\mu}_{R,\mu}\right]}{\exp\left(\frac{k_0}{T}\right)-1},
\end{eqnarray}
where the corresponding imaginary part is given in Ref.~\cite{Wang2020}.

\begin{figure}[htb]
\centering
\subfigure[]{\includegraphics[width=0.45\textwidth]{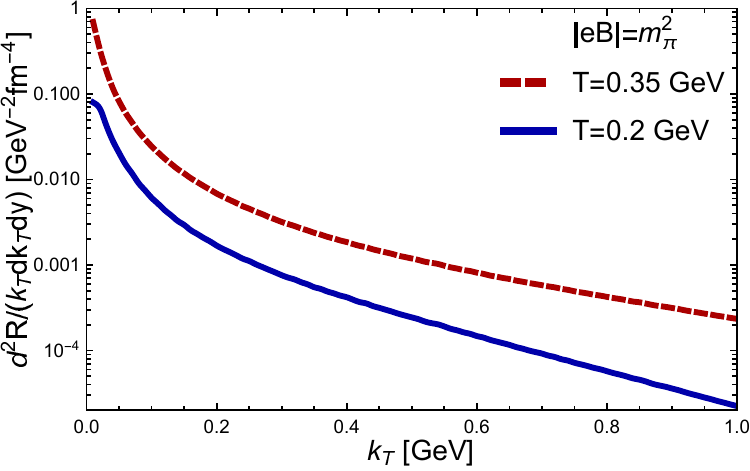}}
  \hspace{0.01\textwidth}
\subfigure[]{\includegraphics[width=0.45\textwidth]{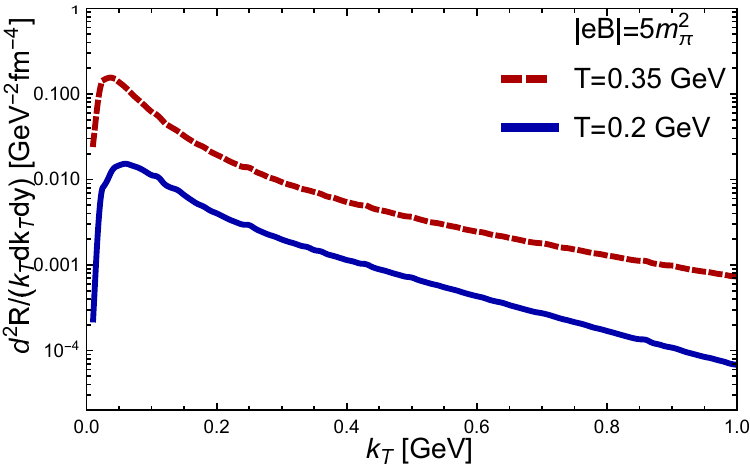}}
\caption{The integrated photon production rate as a function of the transverse momentum $k_T$ for $|eB|=m_\pi^2$ (a) and $|eB|=5m_\pi^2$ (b). The different lines represent results for two different temperatures, i.e., $T=200~\mbox{MeV}$ (blue solid line) and $T=350~\mbox{MeV}$ (red dashed line).}
\label{fig:ProdRateInt}
\end{figure}
The nonmonotonic dependence of the integrated photon production rate on the transverse momentum is shown in Fig.~\ref{fig:ProdRateInt} for 
\(|eB|=m_\pi^2\) and \(|eB|=5m_\pi^2\). For both magnetic fields, the higher temperature leads to a systematically larger rate, reflecting the enhanced thermal population of quark states. In the stronger-field case, well-resolved peaks appear at small \(k_T\), whereas the corresponding structures are less visible for the weaker field. This indicates that the magnetic field not only changes the overall photon yield, but also induces discrete transition structures associated with the Landau-level spectrum.

\section{Photon anisotropy and polarization}
While the previous section considered the azimuthally integrated photon production rate, the magnetic field can also affect the angular distribution of emitted photons. Since the external field selects a preferred spatial direction, the photon spectrum becomes anisotropic. To quantify this angular structure, it is convenient to define the Fourier coefficients $v_n$ of the photon spectrum as follows:
\begin{equation}
v_n(k_T) = \frac{1}{\mathcal{R}_0}  \int_0^{2\pi} \frac{d^3 R}{k_{T} d k_T d\phi dy}  \cos(n \phi) d \phi ,
\label{def:vn-photon}
\end{equation}
where the normalization factor is determined by integrating the emission rate over the azimuthal angle $\phi$, i.e.,
\begin{equation}
\mathcal{R}_0 = \frac{d^2 R}{k_{T} d k_T  dy} = \int_0^{2\pi} \frac{d^3 R}{k_{T} d k_T  dy d\phi} d  \phi .
\end{equation}

\begin{figure}[htb]
\centering
\subfigure[]{\includegraphics[width=0.45\textwidth]{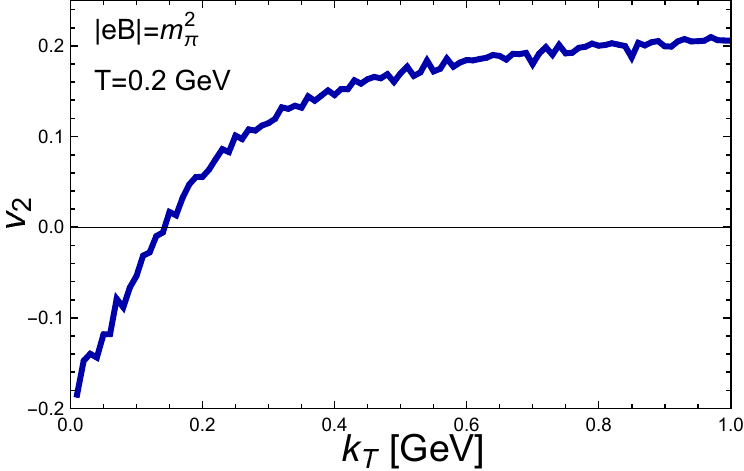}}
  \hspace{0.01\textwidth}
\subfigure[]{\includegraphics[width=0.45\textwidth]{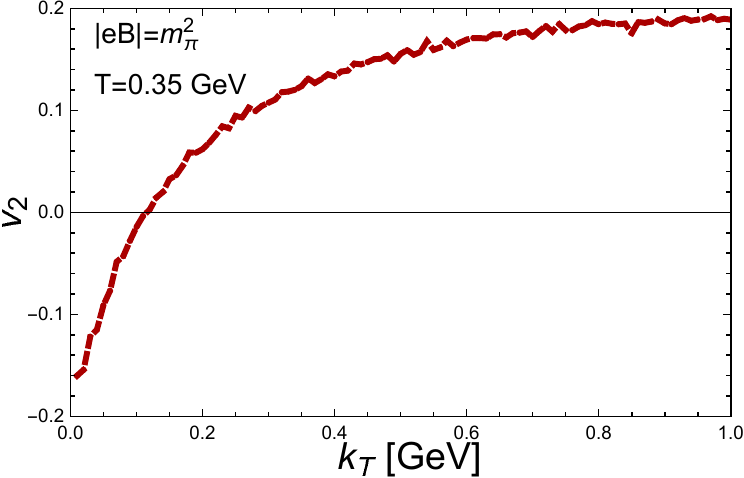}}
\caption{Ellipticity of the photon production as a function of the transverse momentum $k_T$ for $|eB|=m_\pi^2$ and 
two different temperatures: $T=200~\mbox{MeV}$ (a) and $T=350~\mbox{MeV}$ (b).}
\label{fig:B1:v2}
\end{figure}

\begin{figure}[htb]
\centering
\subfigure[]{\includegraphics[width=0.45\textwidth]{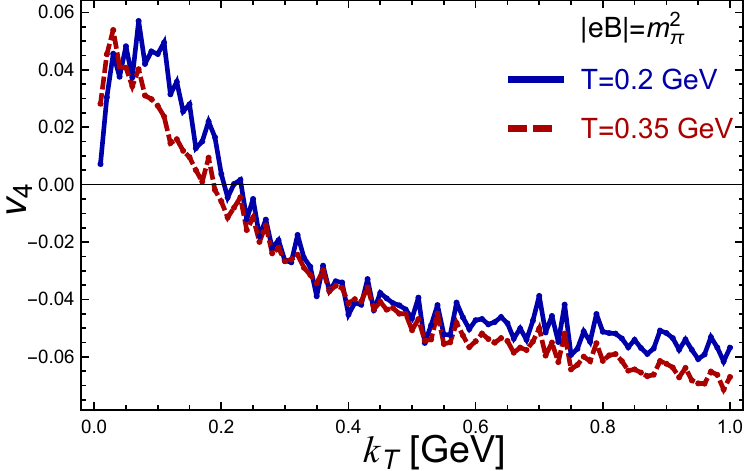}}
  \hspace{0.03\textwidth}
\subfigure[]{\includegraphics[width=0.45\textwidth]{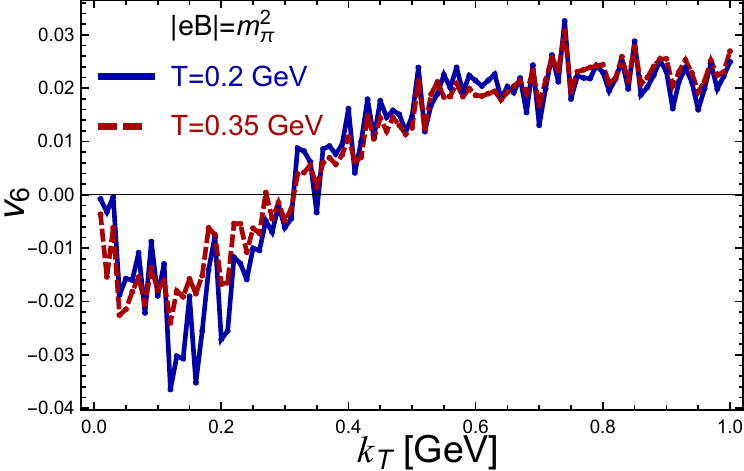}}
\caption{Anisotropic coefficient $v_4$ and $v_6$ for the photon emission as a function of the transverse momentum $k_T$ under $|eB| = m_\pi^2$
for two different temperatures, $T=0.2~\mbox{GeV}$ (blue line) and $T=0.35~\mbox{GeV}$ (red line).}
\label{fig:photon_v46}
\end{figure}

The resulting coefficients \(v_2\), \(v_4\), and \(v_6\) are shown in Figs.~\ref{fig:B1:v2} and \ref{fig:photon_v46}, exhibiting a nontrivial \(k_T\) dependence with both the sign and magnitude varying with temperature and magnetic-field strength. In particular, the even harmonics show opposite sign patterns in the small- and large-\(k_T\) regions. For \(k_T \lesssim \sqrt{|eB|}\), one finds \(v_2<0\), \(v_4>0\), and \(v_6<0\), whereas the signs are reversed for \(k_T \gtrsim \sqrt{|eB|}\). The magnitudes decrease with increasing harmonic order, approximately following a \(1/n^2\) hierarchy. 

These results suggest that the leading even-order anisotropy coefficients \(v_n\) may serve as characteristic signatures of magnetic-field-induced photon emission in a hot QGP. In realistic heavy-ion collisions, however, such a signal would be difficult to isolate because of large backgrounds and the convolution with other sources of anisotropy, such as hydrodynamic flow and initial-state fluctuations. Furthermore, a major theoretical and experimental challenge stems from the highly transient nature of the magnetic fields generated in heavy-ion collisions, which decay rapidly during the early stages of the system's evolution. Therefore, the observable photon anisotropy may depend sensitively on the lifetime of the magnetic field and on the early-time evolution of the QGP. Nevertheless, the present calculation provides a useful theoretical baseline for the intrinsic photon anisotropy generated by the magnetic field. Future improvements in experimental measurements, collision simulations, and data analysis may make it possible to further test these predictions.

\section{Effects of chemical potentials and chiral chemical potentials}
Beyond modifying the total yield and angular anisotropy, a magnetized plasma can also affect the polarization structure of emitted photons, especially in the presence of finite quark-number or chiral charge densities. Using the completeness relation for the photon polarization vectors, we decompose the photon emission rate into different polarization components. 
\begin{equation}
	g^{\mu \nu} = \epsilon_{0}^\mu  \epsilon_{0}^{\nu*} - \epsilon_{+}^{\mu}  \epsilon_{+}^{\nu*} -  \epsilon_{-}^\mu  \epsilon_{-}^{\nu*} - \epsilon_{\parallel}^\mu  \epsilon_{\parallel}^{\nu*} ,
	\label{decompose}
\end{equation}
where $\epsilon_{r}^{\mu}$ are photon polarization vectors, i.e., $\epsilon_{0}^\mu  = (1,0,0,0)$, $\epsilon_{\parallel}^\mu=(0,0,0,1)$,  $\epsilon_{\pm}^{\mu}=(0,1,\pm i, 0)/\sqrt{2}$.

We focus on the emission of circularly polarized photons, represented by the polarization vectors $\epsilon_{\pm}^{\mu}$. For photons emitted parallel or antiparallel to the magnetic field, these states correspond to left- and right-handed circular polarizations, equivalently photons with opposite helicities. When these projectors are contracted with $\Pi_{\mu\nu}(k)$, the photon polarization tensor naturally separates into the following four parts:
\begin{equation}
	g^{\mu \nu}\Pi_{\mu\nu}(k)  \to  \Pi^{(0)}(k) + \Pi^{(+)}(k) + \Pi^{(-)}(k) +\Pi^{(\parallel)}(k) .\
	\label{Pi-split}
\end{equation}
The contributions proportional to \(\mathrm{Im}\,\Pi^{(+)}\) and \(\mathrm{Im}\,\Pi^{(-)}\) define the emission rates of right- and left-handed circularly polarized photons:
\begin{eqnarray}
	{\cal R}^{(+)}_{\rm diff} &\equiv& k^0 \frac{d^3R}{dk_x dk_y dk_z}  
	= - \frac{1}{(2\pi)^3} \frac{\mbox{Im}\left[\Pi^{(+)}(k) \right]}{\exp\left(\frac{k_0}{T}\right)-1},
	\label{diff-rate-plus} \\
	{\cal R}^{(-)}_{\rm diff} &\equiv& k^0 \frac{d^3R}{dk_x dk_y dk_z}  
	=  - \frac{1}{(2\pi)^3} \frac{\mbox{Im}\left[\Pi^{(-)}(k) \right]}{\exp\left(\frac{k_0}{T}\right)-1},
	\label{diff-rate-minus}
\end{eqnarray}

\begin{figure}[htb]
\centering
{\includegraphics[width=0.44\textwidth]{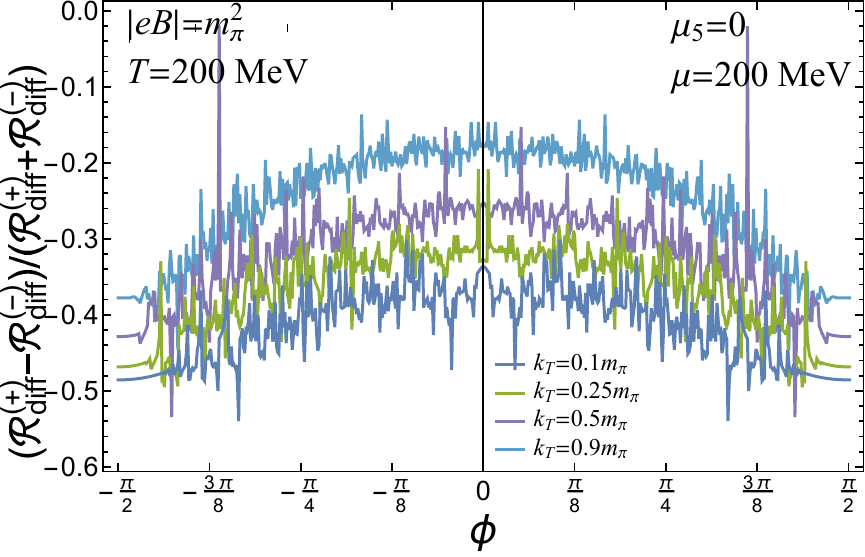}}
  \hspace{0.05\textwidth}
{\includegraphics[width=0.44\textwidth]{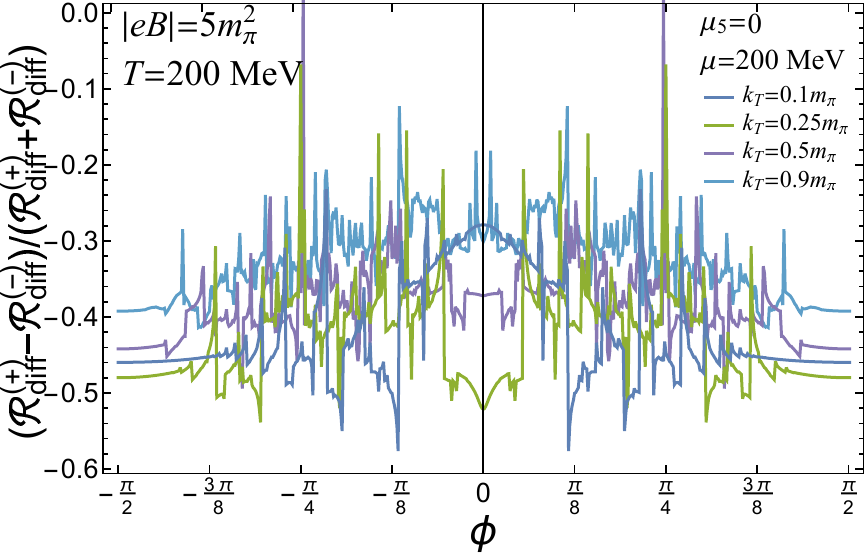}} 
\caption{The degree of circular polarization as a function of the azimuthal angle for $T = 200~\mbox{MeV}$ with two magnetic fields,  $|eB| = m_\pi^2$ (left) and $|eB| = 5m_\pi^2$ (right).}
\label{fig:polarization-degree-mu}
\end{figure}

For finite chemical potential, the emission rates for left-handed and right-handed circular polarizations differ significantly. To quantify this imbalance, we define the degree of circular polarization as
\begin{eqnarray}
	{\cal P}_{\rm circ}  (k_T,\phi, y)= \frac{{\cal R}^{(+)}_{\rm diff} (k_T,\phi, y) - {\cal R}^{(-)}_{\rm diff} (k_T,\phi, y)}{{\cal R}^{(+)}_{\rm diff} (k_T,\phi, y) + {\cal R}^{(-)}_{\rm diff} (k_T,\phi, y)}.
	\label{P-circ}
\end{eqnarray}
The corresponding numerical results are presented in Fig.~\ref{fig:polarization-degree-mu}, which indicate a substantial degree of circular polarization in the photon emission. The sign of ${\cal P}_{\rm circ}  (k_T,\phi, y)$ is determined primarily by a nonzero electrical charge density in the plasma. Indeed, by separating the partial contributions from the up and down quarks, we find that the up quarks emit more left-handed photons (${\cal P}^{\rm (u)}_{\rm circ} <0$), while the down quarks emit more right-handed photons (${\cal P}^{\rm (d)}_{\rm circ} >0$). Since the up quarks have twice the electric charge of the down quarks, their emission rate is higher overall. Consequently, the net circular polarization from QGP is negative, ${\cal P}_{\rm circ} <0$.

\begin{figure}[htb]
\centering
{\includegraphics[width=0.44\textwidth]{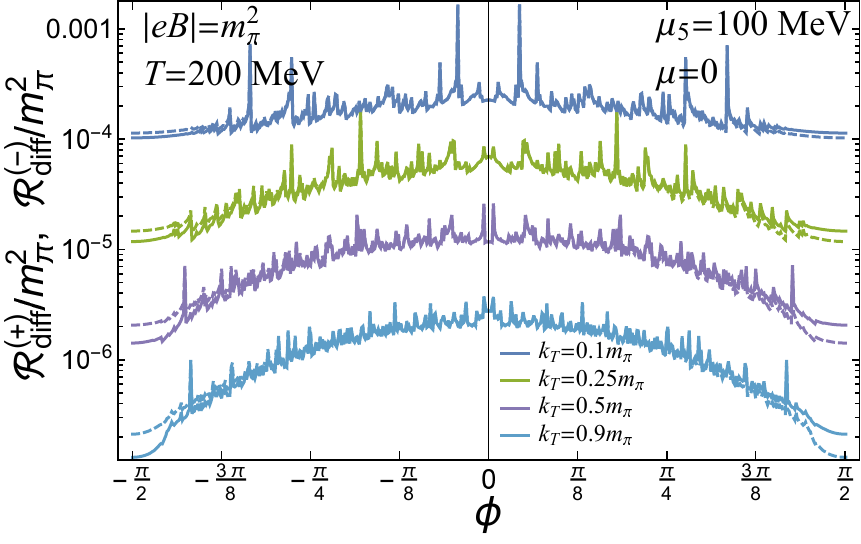}}
  \hspace{0.05\textwidth}
{\includegraphics[width=0.44\textwidth]{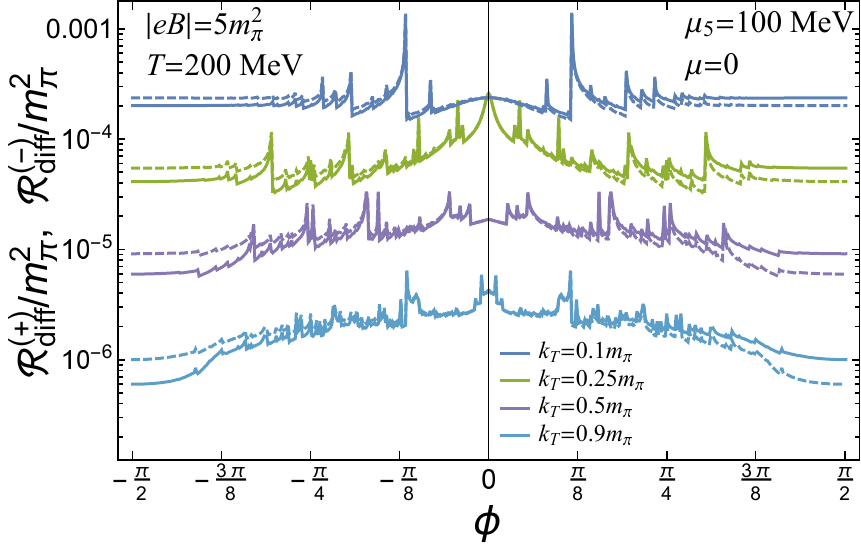}} 
\caption{The angular dependence of right-handed (solid lines) and left-handed (dashed lines) circularly polarized photon emission rates for $T = 200~\mbox{MeV}$ with two magnetic fields,  $|eB| = m_\pi^2$ (left) and $|eB| = 5m_\pi^2$ (right).}
\label{fig:rates-mu5}
\end{figure}

Fig.~\ref{fig:rates-mu5} shows the angular dependence of the left- and right-handed circularly polarized photon emission rates at finite chiral chemical potential. For positive \(\mu_5\), the right-handed component is enhanced along the magnetic-field direction, while the left-handed component is enhanced in the opposite direction. The resulting asymmetry is most pronounced around \(\phi\simeq \pi/2\), and its magnitude increases with transverse momentum. It is also enhanced by a stronger magnetic field, but reduced at higher temperature.

A nonzero quark-number chemical potential also affects the relative strength of the two circular polarizations. For \(\mu>0\), the left-handed component dominates over the right-handed one. This effect may be interpreted as a photon analogue of the Hall effect in a plasma with finite charge density.

When the electric charge density vanishes, \(\mu=0\), the circularly polarized emission is symmetric under reflection in the transverse plane. In contrast, a finite chiral charge density, \(\mu_5\neq0\), breaks this symmetry. For \(\mu_5>0\), the emission rate of right-handed photons is higher in the upper hemisphere, while the left-handed component dominates in the lower hemisphere, where the two hemispheres are defined with respect to the magnetic-field direction. The pattern is reversed when the sign of \(\mu_5\) changes.

These results suggest that circular polarization observables may provide characteristic signatures of finite electric and chiral charge densities in a strongly magnetized plasma. More broadly, the sensitivity of circularly polarized photon emission to $\mu$ and $\mu_5$ suggests a possible connection with anomalous transport phenomena. Specifically, a finite chiral charge density $\mu_5$ under a strong magnetic field is intrinsically tied to the Chiral Magnetic Effect (CME), whereas a finite baryon chemical potential $\mu$ drives the Chiral Separation Effect (CSE). While the present work does not provide a direct hydrodynamic or transport simulation of these macroscopic currents, our results demonstrate that circularly polarized photon emission observables can serve as sensitive, complementary, and non-destructive probes of the local electric and chiral charge densities in a magnetized QGP. For a comprehensive and recent review of strongly interacting matter in extreme magnetic fields and these related anomalous transport mechanisms, we refer the reader to Ref.~\cite{Adhikari:2024bfa}.

\section{Summary and perspective}
In this contribution, we presented several aspects of thermal photon production in magnetized QCD matter in connection with relativistic heavy-ion collisions. We studied photon emission from a magnetized quark-gluon plasma, including anisotropy and the effects of finite chemical potential and chiral chemical potential. In the future, we will extend this study to more realistic phenomenological analyses with the space-time evolution of heavy-ion collisions taken into account.


\bibliographystyle{unsrt}
\bibliography{references}

\end{document}